\documentclass[12pt]{article}
\pdfoutput=1
\usepackage{times}
\usepackage{geometry}
\usepackage[utf8]{inputenc}
\usepackage{enumitem,amssymb}
\usepackage{ragged2e}
\newlist{thematic}{itemize}{8}
\setlist[thematic]{label=$\square$}
\usepackage{pifont}

\usepackage{graphics,graphicx}

\usepackage{float}
\usepackage{lineno}
\usepackage{amssymb}
\usepackage[normalem]{ulem}
\usepackage{enumerate} 

\begin{document}
\pagestyle{plain}
\pagenumbering{arabic}

\raggedright
\huge
Astro2020 Science White Paper \linebreak

Studying the magnetized ISM with all-sky polarimetric radio maps \linebreak
\normalsize

\noindent \textbf{Thematic Areas:}  Galaxy Evolution, Multi-Messenger Astronomy and Astrophysics \linebreak
  
\textbf{Principal Author:}
\vspace{0.2cm}
\linebreak
Name:	{\bf Colin J. Lonsdale}
 \linebreak						
Institution:  Haystack Observatory, Massachusetts Institute of Technology
 \linebreak
Email: cjl@haystack.mit.edu
 \linebreak
Phone:  617 715 5575
 \linebreak
 
\textbf{Co-authors:}
\vspace{0.2cm}
\linebreak						
{\bf Elena Orlando} (Kavli Institute for Particle Astrophysics and Cosmology, W.W. Hansen Experimental Physics Laboratory, Stanford University)
 \linebreak						
{\bf Gregg Hallinan} (California Institute of Technology)
 \linebreak						
{\bf Greg Taylor} (University of New Mexico)
 \linebreak						
{\bf Clive Dickinson} (Jodrell Bank Centre for Astrophysics, University of Manchester, UK)
  \linebreak

\justify

\textbf{Abstract:}
\linebreak
Synchrotron radiation from the interstellar medium (ISM) of our galaxy dominates the sky brightness at low radio frequencies, and carries information about relativistic and thermal electron distributions across a range of astrophysical environments.  The polarization of the radiation, as modified by Faraday rotation effects in the ISM, also contains extensive information about the magnetic field.  Comprehensive all-sky broadband mapping of this radiation, when combined with high frequency radio data, gamma ray data, cosmic ray (CR) measurements and sophisticated modeling, can revolutionize our understanding of the ISM and the processes that influence its evolution.

Current widefield imagery of the galactic synchrotron emission is heterogeneous in frequency coverage, sky coverage, angular resolution and calibration accuracy, limiting utility for ISM studies.  A new generation of all-digital low frequency array technologies is opening a path to matched resolution, high fidelity polarimetric imaging across a fully sampled swath of radio frequencies from a few tens to many hundreds of MHz, generating a transformational dataset for a broad range of scientific applications.

\pagebreak




\section{Introduction} 
The focus of this white paper is study of the interstellar medium, in particular employing high-precision, broadband, low frequency, polarization-sensitive radio imaging of the galactic synchrotron emission to better constrain:
\begin{itemize}
	\item The large-scale Galactic magnetic fields, both the turbulent and ordered components, and their orientation and strength throughout entire Galaxy
	\item The Galactic magnetic field properties, as above, in specific dense regions
	\item The low-energy cosmic rays (CR) that are responsible for the radio emission, including their propagation, interactions, and role in the evolution of our galaxy
\end{itemize}

At low radio frequencies, from the ionospheric cutoff at $\sim$10MHz to at least 500 MHz, the sky brightness is dominated by synchrotron radiation from the Milky Way galaxy.  It is due to relativistic electrons and magnetic fields in the interstellar medium (ISM), is optically thin almost everywhere on the sky across this entire frequency range, and has a spectral index of $\sim$-0.6 in flux density per unit sky area ($\sim$-2.6 in brightness temperature).  

From a given volume element of the ISM, the emission is linearly polarized according to the configuration of the magnetic field.  As it propagates through the ISM, it passes through magnetized plasma with both thermal and relativistic electron populations, leading to both Faraday rotation and additional emission along the path length to the observer.  Thus, as seen from Earth the galactic synchrotron emission is strongly concentrated along the plane (optically thin, long path lengths) and highly structured in linear polarization properties (complex distribution of magnetized thermal plasma, superposition of emission from regions with different field configurations).  In addition, certain sightlines may intersect regions of high thermal electron density (e.g. HII regions, the galactic center region), leading to significant free-free absorption at the lower frequencies

The detailed properties of this emission therefore contain extensive information on the properties of the multiphase ISM, with diagnostics of magnetic fields, the relativistic electron population and the thermal electron distribution.  This allows for detailed studies of astrophysical conditions and processes in a wide variety of environments throughout the galaxy, depending on the accuracy, angular resolution and frequency range of the data.  To date the data quality has not been sufficient to perform in-depth analyses that exploit the scientific potential of such measurements, but the future potential, enabled by emerging technology-driven instrumental capabilities, is profound.

\section{Synergies with other ISM studies}
Sophisticated models are necessary for understanding large-scale magnetic fields and CR distribution and propagation, and such models also incorporate constraints from direct CR measurements (note that the radio emission is due to the low energy CR electrons). Observations of the interstellar synchrotron emission in the radio band from a few MHz to tens of GHz have been used to constrain CRs and propagation models by Strong et al. (2011), favoring models with no or low reacceleration. More recently, Orlando \& Strong (2013) used the spectral, spatial and polarization distribution of the synchrotron emission to explore various CR source distributions, CR propagation halo sizes, propagation models (e.g. pure diffusion and diffusive-reacceleration models), and 3D Galactic magnetic field formulations. These models have been used in the context of low frequency foreground component maps and magnetic field studies (Planck Collaboration, 2016a,b). The synergy with observations at microwaves (Planck) is thus very important, and a better determination of the synchrotron emission is fundamental for separation of superposed foregrounds.

Because the CR electrons responsible for the radio emission also generate gamma-ray emission, synergies with gamma-ray telescopes, e.g. with Fermi LAT and possible upcoming missions at MeV are important (see e.g. Orlando 2018, Orlando 2019). Orlando (2018) explored this synergy by jointly studying the radio, microwave, and gamma-ray interstellar emission, obtaining the local interstellar spectrum of electrons independently of direct CR measurements subject to solar modulation.  Studies in gamma rays (e.g. Ackermann et al. 2012; Tibaldo et al. 2015) and radio (Orlando \& Strong 2013; Fornengo et al. 2014) result in discordant confinement volume estimates for the CRs. Indeed there is a degeneracy between the isotropic component and the CR halo confinement. Another important topic to be addressed by accurately calibrated wideband imagery is the currently unknown origin of the large all-sky radio excess recently extensively reviewed (Singal et al. 2018).

In addition to the large-scale synchrotron emission and depending on achievable angular resolution, some dense regions, such as molecular clouds, could provide information on the magnetic fields and the penetration of CRs in the clouds, which affect star formation (e.g. Dickinson et al. 2015, Padovani et al. 2016, 2018).
 
\section{State of the art in radio measurements} 

The current state of the art in characterization of the synchrotron radiation from the ISM is captured in the datasets and imagery available on the LAMBDA website hosted by NASA GSFC.  The sky coverage and angular resolution of low frequency surveys varies widely, and is generally poorer at the lower frequencies due to technical challenges that have only recently started to be overcome.  For decades, by far the highest quality low frequency all-sky image in terms of angular resolution and completeness has been the Haslam et al (1982) 408 MHz map, most recently reprocessed by Remazeilles et al. (2015).  This map has typically formed the backbone of models seeking to integrate data from a broad frequency range into a global sky model (e.g. de Oliveira-Costa et al. 2008, updated by Zheng et al. 2017), from which researchers can estimate the sky brightness at any frequency for a range of applications, including estimating foreground contamination in cosmological studies.  Unfortunately, current maps below 1 GHz generally have poor calibration internally and relative to each other, making precision broadband work difficult.

Particularly at very low frequencies, an important regime where strong spectral curvature due to absorption processes is increasingly common, maps have very low angular resolution, often of several degrees.  Until recently the best available all-sky image below 100 MHz has been that of Guzm{\'a}n et al (2011), with $\sim$4 degree resolution at 45 MHz.

With few exceptions, these wide field mapping studies do not include precision polarimetry.  The strength, E-vector orientation and detailed frequency dependence of the linear polarization are essential ingredients contributing to the scientific potential of the polarized synchrotron radiation, and the complex modification of that polarization by Faraday effects in the ISM.

The state of the art in low frequency all-sky mapping is summarized in Figure 1.  The Haslam and Calabretta images were generated using single dish instruments with many pointings.  The Guzm{\'a}n image was created using dipole-based phased-array transit instruments. By contrast, the Dowell image was generated interferometrically using a modern digital imaging array, the LWA system (Taylor et al. 2012), and is one of a series spanning the 35-80 MHz range, enabling spectral maps to be constructed.  

\begin{figure}[h!]
\centering
\includegraphics[width=1.0\textwidth]{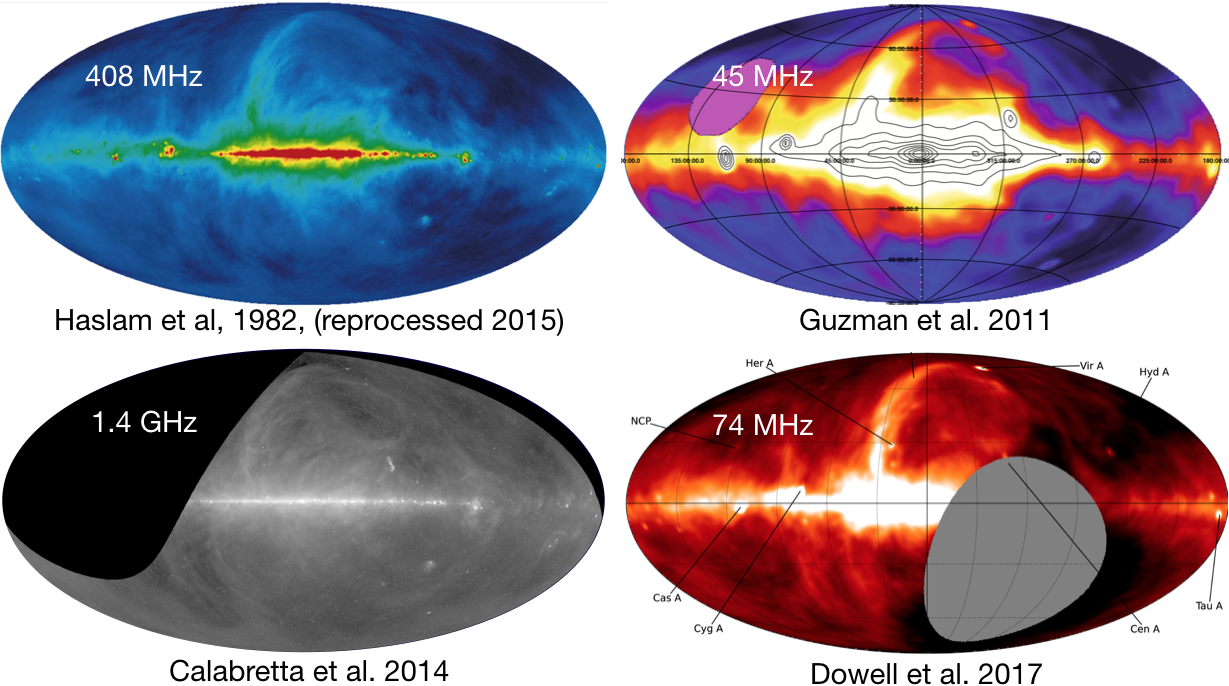}
\caption{\label{fig:fig1} Maps of the continuum sky at different frequencies.  At top left is the heavily used Haslam et al. 408 MHz map with an angular resolution of 51 arcmin.  Top right is the Guzm{\'a}n et al 45 MHz image at $\sim$4 degree resolution.  Lower left is the Calabretta et al. 1.4 GHz image at $\sim$14 arcmin.  The lower right image, by Dowell et al., is an interferometrically reconstructed map using the LWA, at 74 MHz with a resolution of 2.2 degrees}
\end{figure}

\begin{figure}[h!]
\centering
\includegraphics[width=0.9\textwidth]{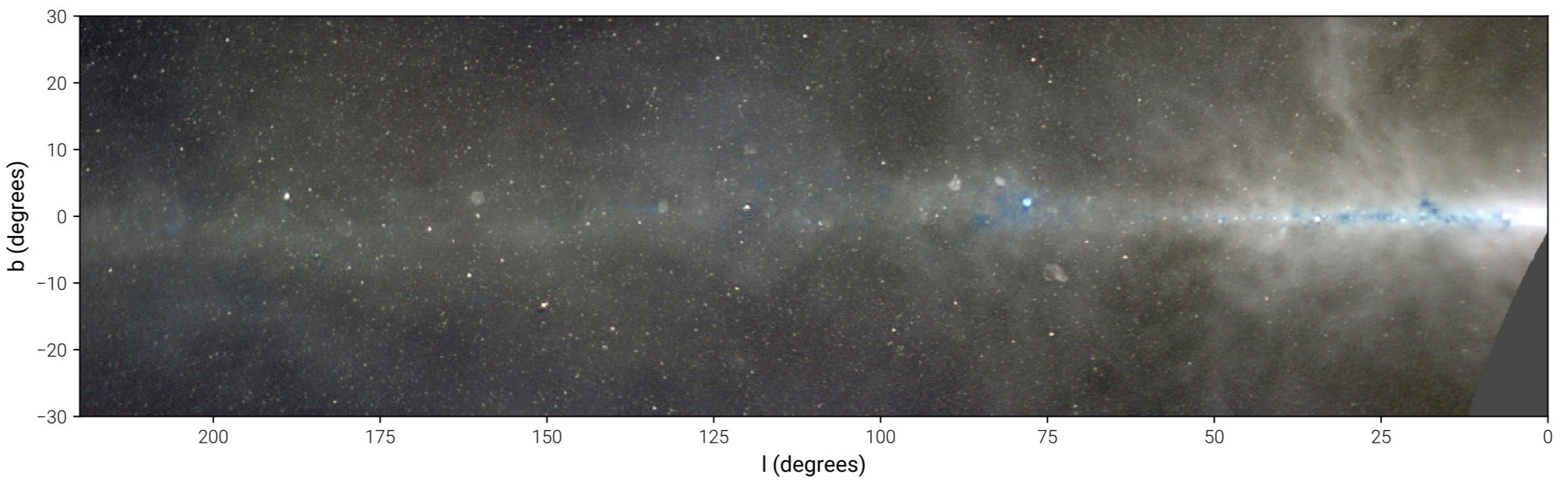}
\caption{\label{fig:fig2} Zoomed in section of the galactic plane from the multifrequency image of Eastwood et al. (2018), showing emission at 15 arcmin resolution from the LWA-OVRO array (Hallinan et al. in preparation).  Blue colors indicate flatter spectra between 36, 52 and 73 MHz.  This image further illustrates the potential of modern digital arrays for this work}
\end{figure}


\section{Future prospects and opportunities}


The broadband synchrotron spectrum reflects the underlying electron energy spectrum, which in turn is determined by electron diffusion and reacceleration processes (e.g. shocks due to supernovae and other phenomena) and electron energy losses (e.g. adiabatic and radiative).   Free-free absorption manifests as flattening of the spectrum, and steep turnover at lower frequencies.  The spectral shape of the observed signal will exhibit significant direction-dependent complexity reflecting integrated emission and absorption conditions along each line of sight.  Due to the spectrally smooth nature of the emitting and absorbing mechanisms, the potential for disentangling multiple effects from each other is a strong function of the continuity and width of the spectral range covered at a given angular resolution.  Precise amplitude calibration across the band is also a prerequisite for robust fitting of physically motivated spectral shape parameters.

Precise maps of the linear polarization of the emission will provide powerful diagnostics of both the thermal magnetized plasma and the magnetic field configuration in the ISM.  Two concepts are central to such studies.  The first is rotation measure (RM) synthesis, which takes advantage of the fact that the Faraday rotation due to a discrete screen of magnetized plasma is proportional to wavelength squared.  One can thus take a broadband measurement of polarization position angle, Fourier transform it along the lambda squared axis, and separate individual screens in Faraday depth (RM), providing extraordinary 3-dimensional views of Faraday-rotating media given adequate spectral coverage (Schnitzeler \& Lee 2015 and references therein).

The second concept is that of a ``polarization horizon".  Multiple mechanisms can render the net observed linear polarization negligible.  This can be due to differential Faraday rotation between different depths in a uniform emitting region along a single line of sight (``slab" depolarization), spatially adjacent cells that are blended together by finite angular resolution (``beam" depolarization), or the superposition of emission with regions that have different field orientations along the same line of sight.  Depolarization becomes stronger with increasing distance from the observer, and will also become stronger at lower frequencies.  Thus any observed linear polarization will generally originate from closer than some distance, depending on frequency and angular resolution.  This ``horizon" is closer at lower frequencies, and for example is tens of parsecs at 150 MHz and ~0.5 degree angular resolution (Lenc et al. 2016).  The strong frequency dependence means that different frequency ranges will probe different distances, yielding 3-dimensional information.

\begin{figure}[h!]
\centering
\includegraphics[width=0.7\textwidth]{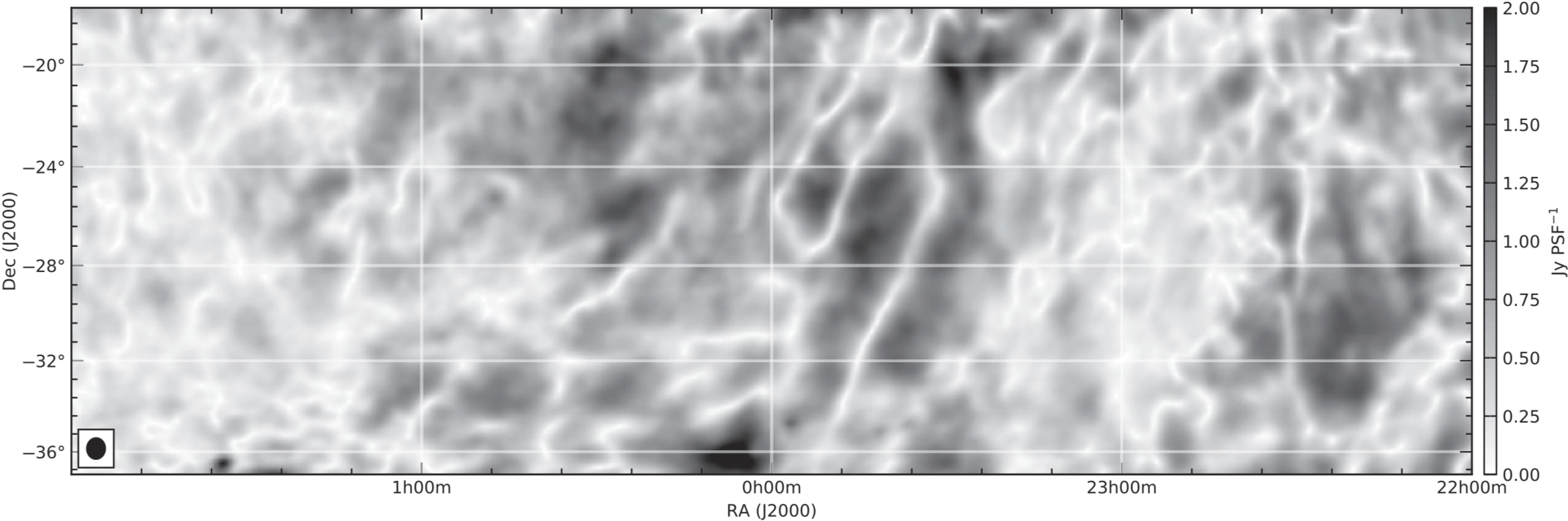}
\caption{\label{fig:fig3} From Lenc et al. (2017), a $\sim$1200 square degree swath of sky as seen in linear polarized intensity by the MWA at 216 MHz with a resolution of $\sim$50 arcmin.  This emission is ubiquitous, strong and easy to detect.  The emission patterns are, however, very complex due to structure in the nearby ISM that causes differential Faraday rotation along different sightlines.}
\end{figure}

In recent years, extensive progress has been made in precise, absolute calibration of low frequency instruments.  The clearest example of this is work for the EDGES experiment, designed to perform precision broadband measurements of the global sky spectrum with the aim of detecting weak spectral features due to redshifted 21cm neutral hydrogen absorption and emission.  By using stringent design approaches, precise measurements of antenna reflection coefficients, detailed electromagnetic modeling of the antenna, and deployment to a remote radio quiet location, the necessary measurement precision for this demanding observation has been demonstrated (Bowman et al. 2018).  As a byproduct of this work, measurements of the spectral index of the galactic synchrotron emission with unprecedented accuracy are possible, with recent results shown in Mozden at al. (2017, 2019), and Figure 4.  The same techniques can be applied to an imaging interferometer, offering the  prospect of exquisitely precise broadband (frequency span of $\sim$10:1 or greater) spectra for each of the $>10^5$ pixels in a sub-degree resolution all-sky image.  Such spectra would support detailed physically motivated spectral decompositions across the whole sky.

\begin{figure}[h!]
\centering
\includegraphics[width=0.4\textwidth]{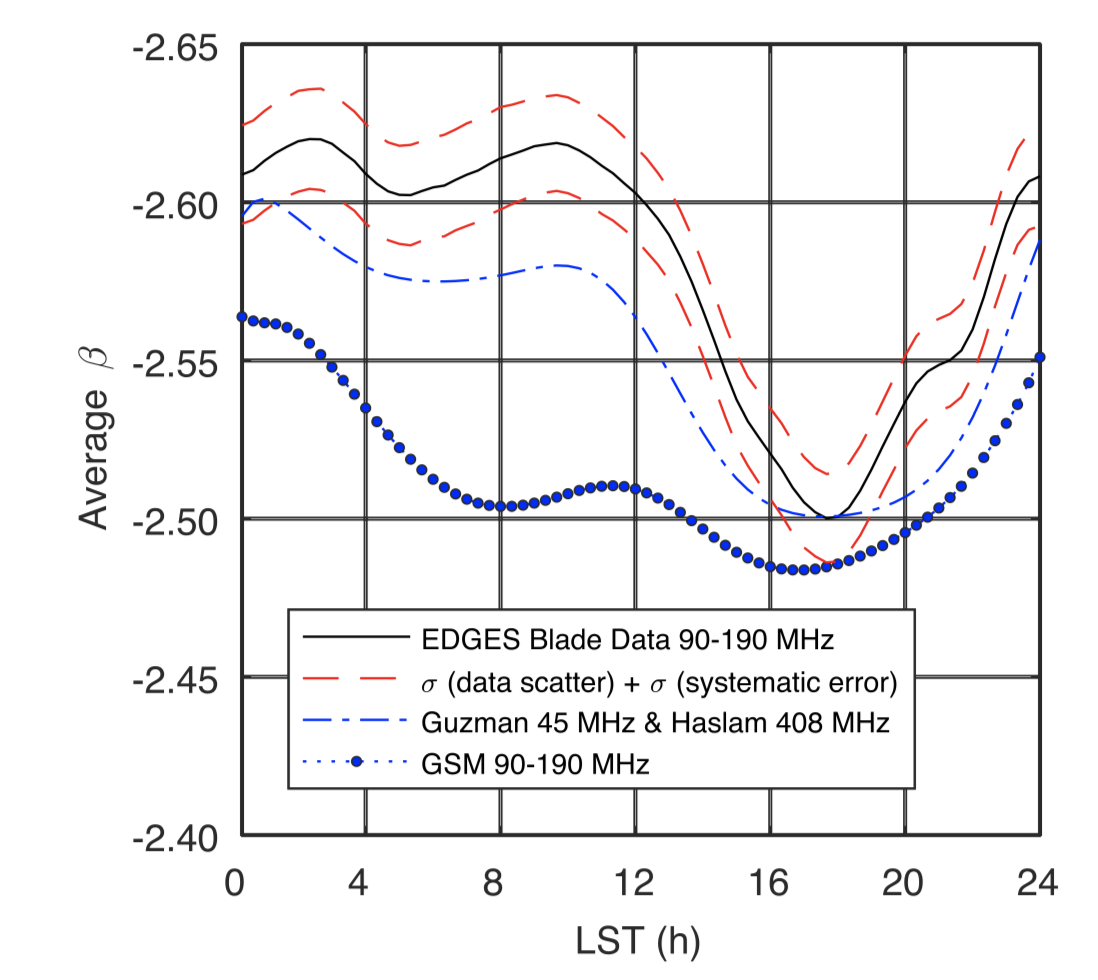}
\caption{\label{fig:fig4} From Mozden et al. (2017), the 90-190 MHz sky temperature spectral index from the EDGES single blade dipole antenna (black line), or as different existing sky models would have been seen by EDGES (blue).  The accuracy is $\sim$0.02 in spectral index.  The spatial resolution is very low, but the absolute spectral measurement precision is extremely high.}
\end{figure}

A similarly rich and spectrally complete dataset featuring linear polarization measurements would support RM synthesis analyses at different distances and frequency ranges to probe the ISM in 3D.  The vision is WMAP/Planck levels of fidelity and sensitivity but at low frequencies with good spectral resolution.

These ambitious goals require a wide range of instrument aperture sizes corresponding to comparable angular resolutions at widely separated frequencies.  The accuracy of pixel-based spectral shape measurements depends critically on image fidelity, which in turn demands very dense Fourier plane sampling (provided by computationally intensive use of a large number of independent antennas).  While no single instrument today incorporates a flexible aperture scale, a decade or more of continuous frequency coverage, and precision absolute calibration, the key enabling technologies have already been demonstrated.  It is time to make a transformational leap forward in studies of the magnetized interstellar medium with revolutionary low radio frequency polarimetric imaging capabilities made possible by recent technological advances.



\pagebreak

\def \aap  {A\&A}
\def \aaps  {A\&AS}
\def \aj  {AJ}
\def \apj  {ApJ}
\def \apjs  {ApJS}
\def \apjl  {ApJL}
\def \aplett  {ApJL}
\def \apss  {AP\&SS}
\def \araa  {ARA\&A}
\def \jcap  {JCAP}
\def \prd {Phy. Rev. D}
\def \ssr {SSRv}
\def \mnras {MNRAS}
\def \nat {Nature}
\def \physrep {Phys. Rept.}
\def \pasj {PASJ}
\def \etal {et~al.~}
\def \rmxaa {RMxAA}
\def \jgr {JGR}
\def \pasp {PASP}


\end{document}